\documentclass[%
showpacs,
twocolumn,
amsmath,amssymb,
aps,
pre,
floatfix,
10pt,
]{revtex4-1}

\usepackage{graphicx}
\usepackage{dcolumn}
\usepackage{bm}

\begin{document}

\title{Phase transition in the majority-vote model on the Archimedean lattices}

\author{Unjong Yu}
 \affiliation{Department of Physics and Photon Science, Gwangju Institute of Science and Technology, Gwangju 61005, South Korea}
 \email[E-mail me at: ]{uyu@gist.ac.kr}

\date{\today}

\begin{abstract}
The majority-vote model with noise was studied on the eleven Archimedean lattices
by the Monte-Carlo method and the finite-size scaling.
The critical noises and the critical exponents were obtained with unprecedented precision.
Contrary to some previous reports, we confirmed that the majority-vote model
on the Archimedean lattices belongs to the two-dimensional Ising universality class.
It was shown that very precise determination of the critical noise is required
to obtain proper values of the critical exponents.
\end{abstract}



\maketitle

\section{Introduction}

The Ising model \cite{Ising25} has played a crucial role in the development of important concepts in
the statistical physics such as the phase transition, critical phenomena, and
the universality class \cite{Brush67,Mccoy12}.
Although it was proposed to explain ferromagnetism,
it can be applied to various phenomena of material systems: 
liquid-gas systems at the critical point, order-disorder transitions in
binary alloy systems, charge-ordering in mixed-valence compounds, etc.
In addition, recently, many kinds of varieties of the
Ising model were proposed to explain social phenomena \cite{stauffer08,sen13}.
For example, spin state of a site in the Ising model may represent
a social opinion or preference of a person, which can be affected by
his or her acquaintances. An Ising-like model is also used to simulate
racial segregation, where different spin states represent people of different races \cite{schelling197}.
In this case, the total number of people of a race is constant
and only the spacial configuration of people can change.

Compared with material systems, the social application of the Ising model
should take into account two points. The first one is that the structure of social
interactions are usually complex networks rather than periodic lattices \cite{Strogatz01}.
It changes the character of the phase transition of the Ising model \cite{Barrat00,Aleksiejuk02}.
The second point is that most of social phenomena are irreversible and out of
equilibrium. Therefore, it is not trivial to apply the equilibrium statistical mechanics
to the nonequilibrium social dynamics. Especially, it is not clear whether the universality
hypothesis, which insists that systems of the same spacial dimension
and the symmetry of the order parameter
share the same critical exponents \cite{Stanley99}, is applied also to nonequilibrium models.

Interestingly, it was proposed that the stochastic nonequilibrium systems with up-down symmetry
belong to the equilibrium Ising universality class in a steady state \cite{Grinstein85}.
If it is true, an Ising-like nonequilibrium model in any kind of two-dimensional (2D) lattice
should have the same critical exponents
as the equilibrium Ising model in the 2D square lattice \cite{Onsager44}.
It was confirmed in a few systems \cite{Tome91,Oliveira93,Ortega98}.
As for the majority-vote model (MVM), which is
one of the well-studied nonequilibrium models in the opinion dynamics with up-down symmetry,
it was confirmed in the 2D square lattice \cite{Oliveira92,Kwak07}.
However, later a few other works on other 2D lattices
(triangular, honeycomb, kagom\'{e}, maple-leaf, and bounce lattices)
reported that critical exponents of the
MVM are different from those of the Ising model \cite{Lima06,Santos11}.
More recently, it was argued again that the MVM on the triangular and honeycomb
lattices belongs to the Ising universality class \cite{Acuna14}.
In the three-dimensional simple-cubic lattice, there is also controversy 
about the universality class: Ref.~[\onlinecite{Yang08}] reported different
critical exponents of the MVM from the Ising universality class,
but Ref.~[\onlinecite{Acuna12}] argued against that.
Therefore, it is important to conclude about the nature of the
phase transition of the MVM on the regular lattice.

\begin{figure}[t]
\includegraphics[width=8.4cm]{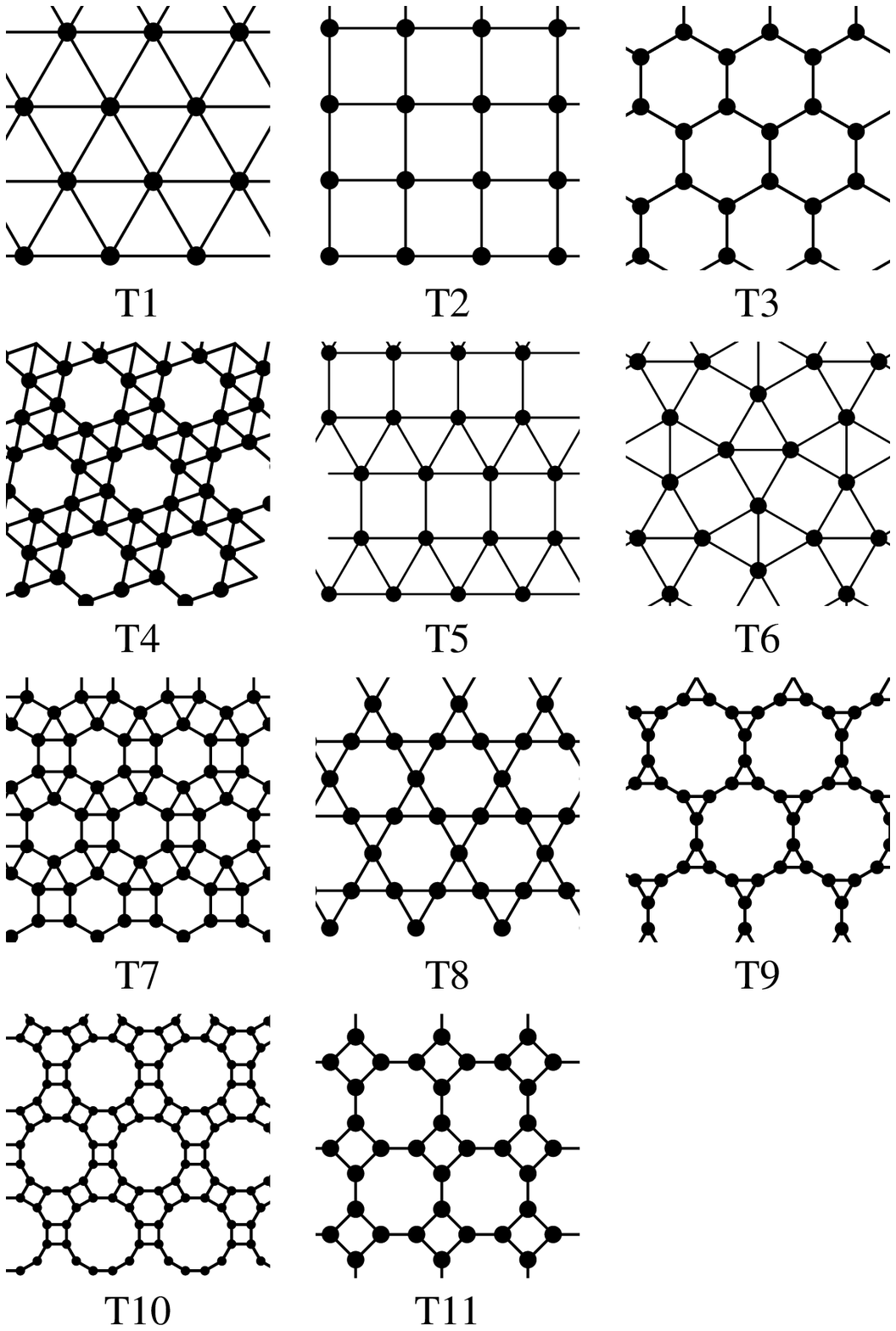}
\caption{The eleven Archimedean lattices.}
\label{Arch}
\end{figure}

In this work, phase transitions and critical phenomena of the MVM
are studied on the eleven Archimedean lattices, which
are 2D lattices by uniform tiling of regular polygons.
It was proved that there exist only eleven Archimedean lattices \cite{Grunbaum87}:
They are listed in Fig.~\ref{Arch} and Table~\ref{Arch_table}.
Due to simplicity and various topologies, they are good bases for systematic
study on 2D systems \cite{Richter04,Yu15b}.
The MVM on six Archimedean lattices has been studied \cite{Oliveira92,Lima06,Kwak07,Santos11,Acuna14},
and the results on the other five lattices are reported for the first time in this paper.
By extensive Monte-Carlo calculations, the critical noise of each
Archimedean lattice was obtained with unprecedented precision,
and it is confirmed that the critical exponents of the MVM
on all of the eleven Archimedean lattices
are same as the 2D Ising universality class.
Possible reasons for the discrepancies of previous works are also discussed.

\begin{table*}[b]
\caption{\label{Arch_table}Name, number of lattice points per basis ($B$),
  coordination number ($z$), number of next-nearest neighbors ($z_2$),
  critical noise ($q_{c}$), critical exponents
  ($\nu$, $\gamma$, and $\beta$), and references for the eleven
  Archimedean lattices within the majority-vote model.
  A next-nearest neighbor is a site with shortest-path-length of two from a given site.
  SHD and CaVO mean square-hexagonal-dodecagonal and CaV$_4$O$_9$, respectively.
  Results of this work are in bold.
}
\begin{ruledtabular}
\begin{tabular}{ccccccccccc}
 & Name & & $B$ & ~$z$~ & ~$z_2$~ & $q_{c}$ & $1/\nu$ & $\gamma/\nu$ & $\beta/\nu$ & Ref.\\
\hline
T1 & Triangular  & ($3^6$)       & 1  & 6 & 12 & 0.114(5) & 1.08(6) & 1.59(5), 1.64(1) & 0.12(4) & \cite{Santos11} \\
                                    & & & & & & 0.1091(1) & 1.01(2) & 1.759(7) & 0.123(2) & \cite{Acuna14} \\
                                    & & & & & & {\bf 0.10910(3)} & {\bf 1.01(5)} & {\bf 1.76(1)} & {\bf 0.125(3)} &  \\ \\
T2 & Square      & ($4^4$)       & 1  & 4 & 8 & 0.075(1) & 0.99(5) & 1.70(8) & 0.125(5) & \cite{Oliveira92} \\
                                    & & & & & & 0.075(1) & 0.98(3) & 1.78(5) & 0.120(5) & \cite{Kwak07} \\
                                    & & & & & & {\bf 0.07518(3)} & {\bf 0.99(3)} & {\bf 1.75(1)} & {\bf 0.123(3)} &  \\ \\
T3 & Honeycomb   & ($6^3$)       & 2  & 3 & 6 & 0.089(5) & 0.87(5) & 1.64(5), 1.66(8) & 0.15(5) & \cite{Santos11} \\
                                    & & & & & & 0.0639(1) & 1.01(2) & 1.755(8) & 0.123(2) & \cite{Acuna14} \\
                                    & & & & & & {\bf 0.06400(3)} & {\bf 1.01(4)} & {\bf 1.75(1)} & {\bf 0.122(3)} &  \\ \\
T4 & Maple leaf  & ($3^4,6$)     & 6  & 5 & 9 & 0.134(3) & 0.98(10) & 1.632(35) & 0.114(3) & \cite{Lima06} \\
                                    & & & & & & {\bf 0.09670(3)} & {\bf 0.99(6)} & {\bf 1.75(1)} & {\bf 0.124(3)} &  \\ \\
T5 & Trellis     & ($3^3,4^2$)   & 2  & 5 & 10 & {\bf 0.10266(3)} & {\bf 1.01(6)} & {\bf 1.75(1)} & {\bf 0.125(3)} &  \\ \\
T6 & Shastry-Sutherland & ($3^2,4,3,4$) & 4  & 5 & 11 & {\bf 0.10930(3)} & {\bf 1.04(6)} & {\bf 1.75(1)} & {\bf 0.125(3)} &  \\ \\
T7 & Bounce      & ($3,4,6,4$)   & 6  & 4 & 8 & 0.091(2) & 0.872(85) & 1.596(54) & 0.103(6) & \cite{Lima06} \\
                                    & & & & & & {\bf 0.05940(3)} & {\bf 0.99(4)} & {\bf 1.75(1)} & {\bf 0.123(3)} &  \\ \\
T8 & Kagom\'e    & ($3,6,3,6$)   & 3  & 4 & 8 & 0.078(2) & 0.86(6) & 1.64(3), 1.62(5) & 0.14(3) & \cite{Santos11} \\
                                    & & & & & & {\bf 0.06192(3)} & {\bf 1.03(5)} & {\bf 1.75(1)} & {\bf 0.125(3)} &  \\ \\
T9 & Star        & ($3,12^2$)    & 6  & 3 & 4 & {\bf 0.00435(4)} & {\bf 1.02(6)} & {\bf 1.73(3)} & {\bf 0.124(5)} &  \\ \\
T10& SHD         & ($4,6,12$)    & 12 & 3 & 5 & {\bf 0.04282(3)} & {\bf 1.02(8)} & {\bf 1.75(1)} & {\bf 0.125(3)} &  \\ \\
T11& CaVO        & ($4,8^2$)     & 4  & 3 & 5 & {\bf 0.04925(3)} & {\bf 1.00(8)} & {\bf 1.75(1)} & {\bf 0.124(3)} &  \\ \\
\multicolumn{3}{c}{2D Ising model (exact)} & & & &               & 1             & 1.75          & 0.125         & \cite{Onsager44} \\
\end{tabular}
\end{ruledtabular}
\end{table*}

\section{Model and methods}
The MVM used in this work is defined
by the following spin flip probability \cite{Oliveira92}:
\begin{eqnarray}
w(\sigma_i \rightarrow -\sigma_i ) = \frac{1}{2} \left[ 1-(1-2q)\, \sigma_i \,
       \mathrm{sgn}\!\!\left( \sum_{j=\mathrm{NN}(i)} \sigma_{j} \right) \right] ~,
       \label{update_rule}
\end{eqnarray}
where the spin $\sigma_i$ at site $i$ can have only $\pm 1$ and the summation is over
nearest neighbors (NN) of site $i$. The function $\mathrm{sgn}(x)$ is the sign function,
which gives $+1$, $-1$, and zero for positive, negative, and zero $x$, respectively.
This update rule is a sort of death-birth dynamics, where one site is chosen
to forget its spin state and its nearest neighbors determine
a new spin state of the site \cite{Allen14}.
In the MVM, it follows the spin of the majority with probability $(1-q)$
and that of the minority with probability $q$. If the two kinds of neighbors
tie, the spin is determined at random. The parameter $q$ is called the noise.
For small $q$, one of the two spin states will prevail the system, and the
system will fluctuate randomly without order for $q$ close to half. It was
shown that there exists a continuous phase transition at the critical 
noise $q_c$ between the two phases \cite{Oliveira92}. 
When $q$ is larger than half,
the site prefers to choose the minority spin of nearest neighbors
and another phase transition can exist \cite{Sastre16}.

In the case of equilibrium spin models, any Monte-Carlo dynamics
that satisfies the detailed balance and the ergodicity gives
equivalent results if the simulation is properly performed.
However, nonequilibrium models depend on details of the update rule,
and so cluster-update algorithms \cite{Swendsen87},
which mitigate the critical slowing down, cannot be used.
In addition, since there is no concept of energy and Boltzmann distribution \cite{Kwak07},
extended ensemble methods \cite{Iba01}
also fail. Therefore, we performed the simulation
directly using the update rule of Eq.~(\ref{update_rule}).
Most of the results were obtained by decreasing the noise $q$ very slowly
from $q=1/2$ with random initial spin state, and the absence of hysteresis
was verified. At each temperature, $5\times 10^{6}$ warming-up Monte-Carlo steps
were followed by $10^{9}$ steps for measurement near the critical noise.
Each site has a chance to flip its spin one time on average per one Monte-Carlo step.
All the calculations were performed on parallelograms of size $N=B\times L_0 \times L_0$,
with the number of sites per basis $B$ and the number of bases in one direction $L_0$.
The periodic boundary condition was used.

At each temperature, the magnetization $m$, the magnetic susceptibility $\chi$,
and the fourth-order Binder cumulant $U$ are calculated as follows \cite{Oliveira92}.
\begin{eqnarray}
m=\frac{1}{N}\left|\sum_{i=1}^{N}\sigma_i\right| ~, \\
\chi=N\left(\langle m^2\rangle - \langle m \rangle^2\right) ~,  \label{Eq_chi}\\
U=1- \frac{\left\langle m^4 \right\rangle}{3 \left\langle m^2 \right\rangle ^2} ~. \label{Eq_Binder}
\end{eqnarray}
In the definition of the susceptibility $\chi$, sometimes a function of $q$
is multiplied \cite{Acuna12,Acuna14,Sastre16,Kwak07,Yang08},
but critical exponents are not affected by that
because the functions do not show critical behavior at $q_c$.
Since energy and specific heat are not defined,
only three critical exponents ($\beta$, $\gamma$, and $\nu$) are defined \cite{Fisher72,FSS}: 
\begin{eqnarray}
m(L,q)=L^{-\beta/\nu} \tilde{m}[L^{1/\nu} (q-q_c)] , \label{Eq_scale_m} \\
\chi(L,q)=L^{\gamma/\nu} \tilde{\chi}[L^{1/\nu} (q-q_c)] , \label{Eq_scale_chi}\\
U(L,q)=\tilde{U}[L^{1/\nu} (q-q_c)] , \label{Eq_scale_U} \\
\frac{dU(L,q)}{dq}= L^{1/\nu} \tilde{U}'[L^{1/\nu} (q-q_c)] , \label{Eq_scale_dUdq}
\end{eqnarray}
where $L=\sqrt{N}$ is the linear size of the cluster and
$\tilde{m}(x)$, $\tilde{\chi}(x)$, and $\tilde{U}(x)$ are scaling functions.
The critical noise $q_c$ is determined first, and the
critical exponents are obtained from the 
physical quantities ($m$, $\chi$, and $dU/dq$) calculated at $q_c$
\cite{Oliveira92,Lima06,Kwak07,Yang08,Santos11,Acuna12,Acuna14}.

\section{Results and Discussion}

\begin{figure}[tb]
\includegraphics[width=8.6cm]{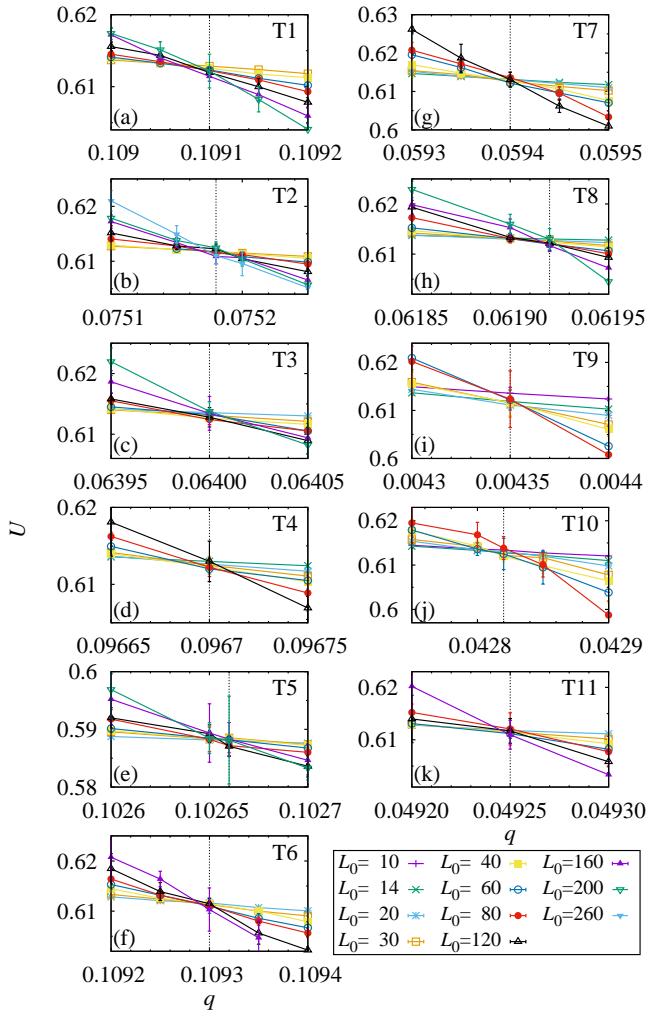}
\caption{(Color online) Binder cumulant ($U$) as a function of noise ($q$)
for the majority-vote model on the eleven Archimedean lattices with various linear sizes ($L$).
The critical noises ($q_c$) obtained by the crossing of the Binder cumulant are shown in vertical
dashed lines.}
\label{Binder}
\end{figure}

\begin{figure}[tb]
\includegraphics[width=8.6cm]{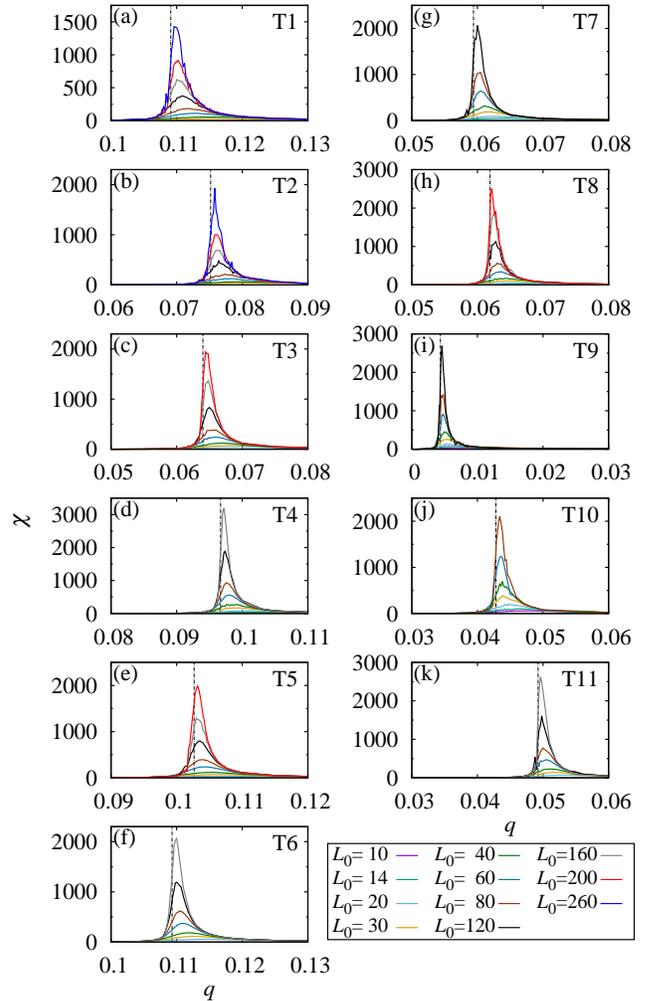}
\caption{(Color online) Magnetic susceptibility ($\chi$), which is defined in Eq.~(\protect\ref{Eq_chi}),
as a function of noise ($q$) for the majority-vote model on the eleven Archimedean lattices
with various linear sizes ($L$).
The critical noises ($q_c$) obtained in Fig.~\protect\ref{Binder}
are represented by vertical solid lines.}
\label{chi}
\end{figure}

The critical noise $q_c$ can be determined by the maximum of the magnetic susceptibility $\chi$
or the derivative of magnetization $dm/dq$. The peak position $q_c(L)$ depends on the linear cluster size $L$
by $q_c(L) = q_c(L=\infty) + \lambda L^{-1/\nu}$, where the parameter $\lambda$ depends
on the lattice type and the physical quantity measured \cite{FSS}. However, it is very difficult
to locate the peak position precisely without extended ensemble methods \cite{Iba01}.
Alternatively, the critical noise can be found by the crossing of the fourth-order Binder
cumulant, because the Binder cumulant becomes independent of the cluster size 
at $q=q_c$ by Eq.~(\ref{Eq_scale_U}),
ignoring the correction critical exponent ($\omega$) \cite{Yu15a}.
The error from the ignorance of the correction critical exponent is
less than statistical error in this calculation.
Figure~\ref{Binder} shows the Binder cumulant
close to the critical point, and the critical noises
obtained by this method are listed in Table~\ref{Arch_table}.
The magnetic susceptibility data in Fig.~\ref{chi} support the results:
The maximum susceptibility position approaches $q_c$ with increasing
the cluster size and it is very close to $q_c$ for large clusters.
As is expected, a lattice with more nearest neighbors tends to
have higher critical noise, but differently from the Ising model,
there are some exceptions \cite{Acuna14}.
When the number of nearest neighbor is same,
a lattice with more number of next-nearest neighbors
has larger critical noise. (Here, a next-nearest neighbor means
a site with shortest-path-length of two from a given site.)

Our results of the critical noise
are consistent with Ref.~\cite{Oliveira92,Kwak07,Acuna14},
but those of Ref.~\cite{Lima06,Santos11} are 
much larger than ours out of error bars.
They used smaller number of Monte-Carlo steps ($2\times 10^5$)
for warming-up at each temperature, but we confirmed it is enough for moderate
size clusters ($N \lesssim 60000$). We performed the simulation
with the same method explained in their papers,
but failed to reproduce their results. 
Therefore, it would be reasonable to guess tentatively
that there is an error in their code.
It would be very difficult to find out the error, if there exists,
without the code they used.
However, the implementation of the lattice seems to be correct, 
because one of the authors could reproduce the exact values
of Curie temperature for the ferromagnetic Ising model \cite{Codello10}
within 1\% in the maple-leaf and bounce lattices \cite{malarz05}.
Therefore, we guess something might be problematic
in the majority-vote dynamics or in calculations of the physical quantities.
Note that similar suspicion was raised in Ref.~[\onlinecite{Malakis12}]
in relation to the spin-1 Ising model.
Wrong value of the critical noise must affect critical exponents seriously.

\begin{figure}[tb!]
\includegraphics[width=8.6cm]{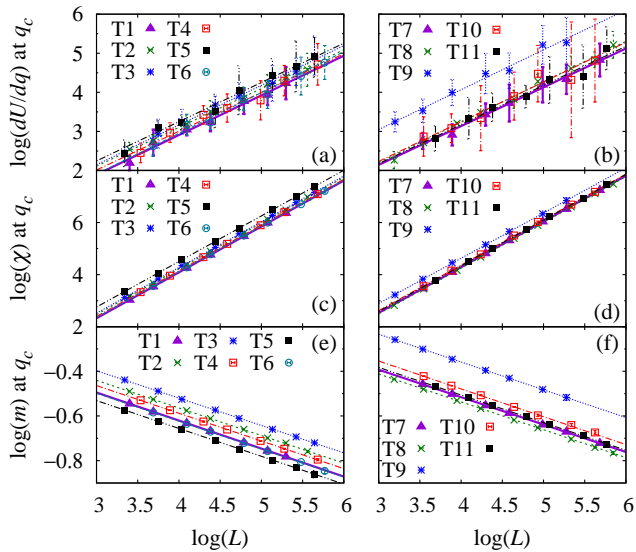}
\caption{(Color online) Derivative of Binder cumulant (top),
magnetic susceptibility (middle), and
magnetization (bottom) at the critical noise $q_c$
for the majority-vote model on the eleven Archimedean lattices
as a function of linear cluster size $L=\sqrt{N}$ in log-log scale.
The straight lines are linear fit, whose slopes represent 
critical exponents ($1/\nu$, $\gamma/\nu$, and $\beta/\nu$ from top to bottom).
They are listed in Table~\protect\ref{Arch_table}.
}
\label{exponent}
\end{figure}

\begin{figure}[tb!]
\includegraphics[width=8.0cm]{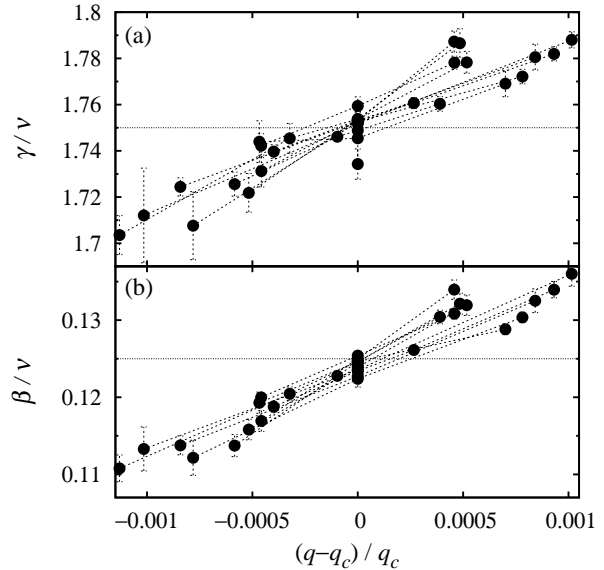}
\caption{Critical exponents $\gamma/\nu$ and $\beta/\nu$
as a function of noise $q$ near the critical noise $q_c$.
These are obtained from fitting of magnetic susceptibility
and magnetization at noise $q$.
The error bars here do not take into account the uncertainty of $q_c$.
The critical exponents of the two-dimensional
Ising universality class are denoted by dashed horizontal lines
($\gamma/\nu=1.75$ and $\beta/\nu=1.25$).
}
\label{exponent_q}
\end{figure}

There are two ways to calculate critical exponents.
Critical exponents $\nu$ and $\gamma$ can be found
from the maximum values of physical quantities of
Eqs.~(\ref{Eq_scale_chi}) and (\ref{Eq_scale_dUdq}).
This method is not efficient without extended ensemble methods.
The second method is to calculate the physical quantities
at the critical point. After the critical noise is
settled, all the critical exponents can be obtained
from the calculation only at the point.
We used this method and the results are shown in Fig.~\ref{exponent} and Table~\ref{Arch_table}.
The linearity is remarkable. Large error bars for $dU/dq$ are from numerical derivative
of $dU/dq = [U(q+\delta)-U(q-\delta)]/(2\delta)$.
This kind of calculation has substantial error inevitably,
since small $\delta$ is required
to reduce discretization error, but it increases the error in the numerator
at the same time. This problem becomes more serious especially when
the function has statistical error like this work. Therefore,
$1/\nu$ has much larger uncertainty than $\gamma/\nu$ and $\beta/\nu$.

Another important source of error is from inaccurate critical point.
To estimate this kind of error, the critical exponents
$\gamma/\nu$ and $\beta/\nu$ are calculated assuming other values
of critical noise close to the correct critical point.
As shown in Fig.~\ref{exponent_q}, values of critical exponents change
a lot with varying the noise $q$: When $q$ change by 0.1\%, 
$\gamma/\nu$ and $\beta/\nu$ change by about 2\% and 8\%, respectively.
For example, if we judge that $\beta/\nu$ should be obtained
within uncertainty of 8\% to verify the universality class,
uncertainty of 0.1\% is allowed at most for the critical noise.
Since the critical noise in the square lattice is $q_c=0.07518(3)$,
the maximum acceptable uncertainty is about $0.0007$, which is not
satisfied by previous works \cite{Oliveira92,Kwak07}.
Thus, it can be argued that the uncertainty of critical exponents
of all previous works are rather underestimated and their conclusions
based on these calculations should be re-examined.
Without correct and accurate calculation of the critical noise,
accurate calculation of critical exponents is impossible and
the decision about the universality class should be inconclusive.
We estimate this is the source of a controversy
about the universality class in this model.
The uncertainty of the critical exponents of our work in Table~\ref{Arch_table}
includes the effect of uncertainty of the critical noise.
In the case of the critical exponent $1/\nu$,
it has large uncertainty already and it is relatively insensitive
to the value of the critical noise.

\begin{figure}[tb!]
\includegraphics[width=8.4cm]{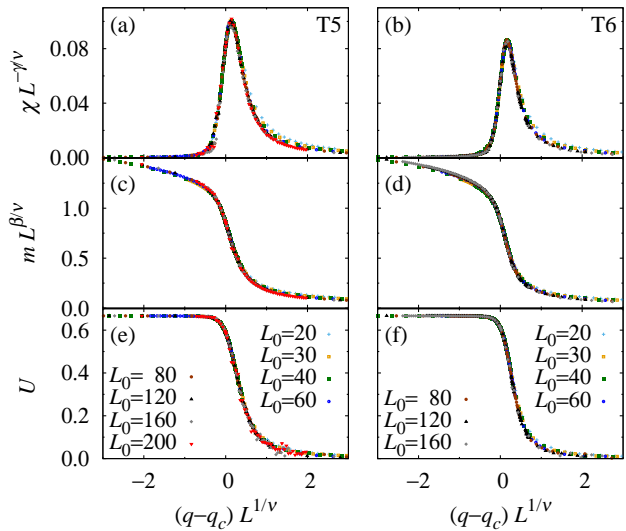}
\caption{Rescaled susceptibility ($\chi L^{-\gamma/\nu}$), 
 magnetization ($m L^{\beta/\nu}$), and Binder cumulant ($U$)
as a function of rescaled noise [$(q-q_c)L^{1/\nu}$] for
the trellis lattice (T5) and the Shastry-Sutherland lattice (T6).
The critical exponents used in this figure are from Table~\protect\ref{Arch_table}.
}
\label{scaling}
\end{figure}

Scaling functions for susceptibility, magnetization,
and Binder cumulant for different system sizes $L_0$
are given in Fig.~\ref{scaling}.
Only two cases are shown, but the other nine lattices
show the same features.
The critical exponents obtained by this work were used.
They all show clear scaling behavior confirming
Eqs.~(\ref{Eq_scale_m}), (\ref{Eq_scale_chi}), and (\ref{Eq_scale_U})
and critical exponents obtained in this work.
We found that variation of critical exponents by a few percent 
does not change the scaling plot noticeably.
Therefore, accurate determination of critical exponents
by scaling plot is very difficult.

\section{Conclusion} 

The MVM was studied on the eleven Archimedean lattices. The critical 
noises were calculated with unprecedented accuracy and the critical
exponents are obtained based on them.
All the critical exponents are same
as those of the 2D Ising model within error bars.
We conclude that the MVM belongs to the Ising universality class
at least within the Archimedean lattices.
We also showed that very accurate determination of the critical
noise is required to calculate the critical exponents properly.

\section*{Acknowledgments}
This work was supported by the GIST Research Institute (GRI) in 2016.

\bibliography{mv1}

\end{document}